\documentclass{article}
\usepackage{spconf,amsmath,graphicx}
\usepackage{amssymb}
\usepackage{pifont}
\newcommand{\xmark}{\ding{55}}

\title{Self-Supervised Learning based Monaural Speech Enhancement with Multi-Task Pre-Training}

\name{Yi Li$^{1}$, Yang Sun$^2$, Syed Mohsen Naqvi$^1$}

\address{\textsuperscript{\rm 1}{ Intelligent Sensing and Communications Research Group, Newcastle University, UK}\\
\textsuperscript{\rm 2}{ Big Data Institute, University of Oxford, UK}}

\begin{document}
%
\maketitle
\begin{abstract}

In self-supervised learning, it is challenging to reduce the gap between the enhancement performance on the estimated and target speech signals with existed pre-tasks. In this paper, we propose a multi-task pre-training method to improve the speech enhancement performance with self-supervised learning. Within the pre-training autoencoder (PAE), only a limited set of clean speech signals are required to learn their latent representations. Meanwhile, to solve the limitation of single pre-task, the proposed masking module exploits the dereverberated mask and estimated ratio mask to denoise the mixture as the second pre-task. Different from the PAE, where the target speech signals are estimated, the downstream task autoencoder (DAE) utilizes a large number of unlabeled and unseen reverberant mixtures to generate the estimated mixtures. The trained DAE is shared by the learned representations and masks. Experimental results on a benchmark dataset demonstrate that the proposed method outperforms the state-of-the-art approaches.

\end{abstract}
\vspace{-0.5em}
\section{Introduction}
\label{sec:intro}
Deep learning techniques have been extensively utilized in speech enhancement for teleconferencing, automatic speech recognition (ASR), and hearing aids \cite{CSA1}\cite{xian}. However, the novel networks are predominantly trained in a supervised mechanism. A vast training set of clean speech signals is required to be well-labelled in the training stage and suffers from drawbacks such as the strong possibility of a mismatch between the training and inference conditions \cite{ssl}\cite{ssl2}. To relax the constraints of supervised learning approaches, self-supervised learning (SSL) based speech enhancement aims to train the model without large labelled datasets to reconstruct the target speech signal from noisy speech. Therefore it becomes highly practical and attractive.

Recently, the SSL techniques have been applied in speech enhancement problem. Wang et al. use an autoencoder to learn a latent representation of clean speech signals \cite{ssl}. However, the pre-training stage only consists of one pre-task which is the mapping of the clean speech spectrogram. Then, Kataria et al. propose a framework called Perceptual Ensemble Regularization Loss (PERL) which shows effectiveness on SSL PASE+ models \cite{ssl3}\cite{PASE}. However, the PERL is limited with the requirement of massive training data.

Followed by our previous work \cite{IET}, to further improve the speech enhancement performance, we introduce both the dereverberation mask (DM) and the estimated ratio mask (ERM) to provide the time-frequency relationships between the clean speech signal and the reverberant mixture. Hence, inspired by \cite{ssl1}, we propose a multi pre-tasks SSL method which only needs a limited set of randomly selected clean speech signals and the corresponding mixture recordings in the pre-training.

\begin{figure*}[htbp!]
\centering
\includegraphics[width=16cm, height=6cm]{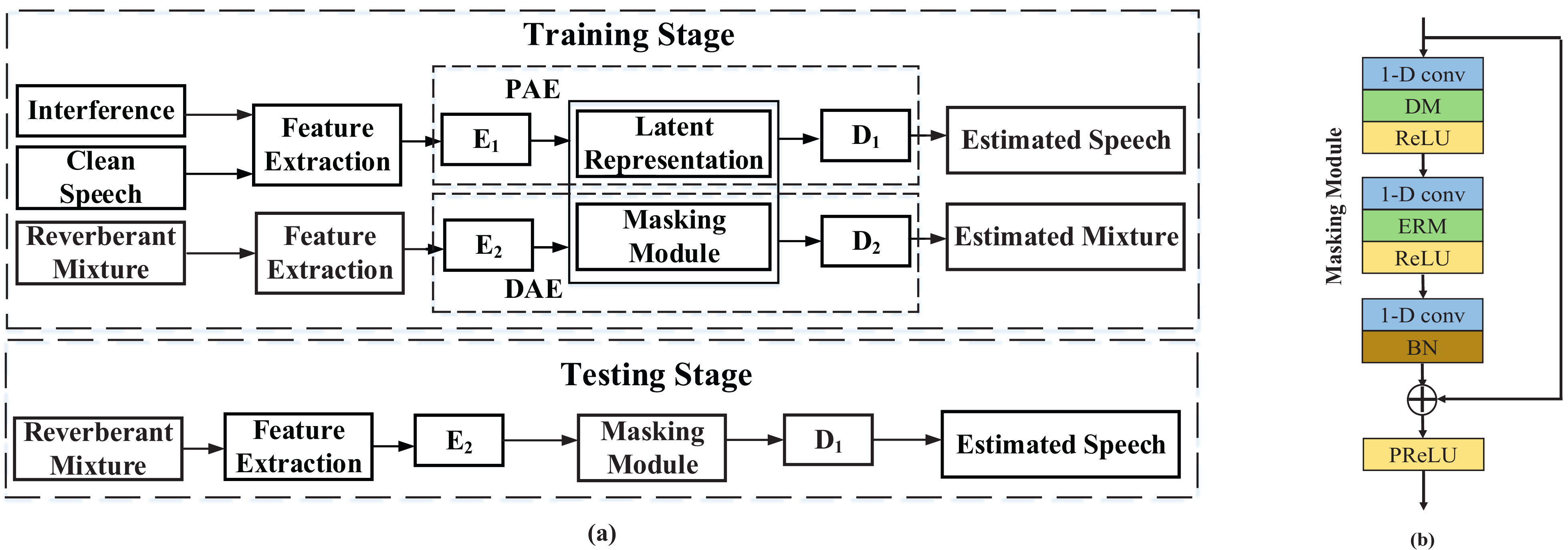}
\caption{The block diagram of the proposed method is shown in (a). The masking module is shown in (b). Features are extracted as the input to the pre-task autoencoder (PAE). The latent representation of clean speech signal is learnt, meanwhile, the target speech signal in the reverberant mixture is estimated in the masking module. The estimated mixture is produced from the downstream task autoencoder (DAE) which shares the learned representation and masks. The enhanced signal is obtained from the output of the decoder in the testing stage.}\centering
\end{figure*}


The contributions of this paper are summarized as follows:

$\bullet $ Multi pre-tasks with self-training are proposed to solve the speech enhancement problem.

$\bullet $ To address the speech enhancement problem with reverberant environment, we apply dereverberation mask in the masking module to dereverberate the mixture.

\section{PROPOSED METHOD}
\label{sec:format}
\subsection{Multi pre-tasks based autoencoders}

The block diagram of the proposed method is shown in Fig. 1 (a). In the training stage, we exploit two variational autoencoders, pre-training autoencoder (PAE) and downstream task autoencoder (DAE), for different tasks. The encoder and decoder of the PAE are denoted as $E_{1}$ and $D_{1}$, respectively. Similarly, we use $E_{2}$ and $D_{2}$ to present the encoder and decoder of the DAE respectively.

The input of the pre-task consists of a limited set of clean speech signals, background noise, and reverberated both speech and noise signals. The mel-frequency cepstral coefficients (MFCC) feature \cite{mfcc} is first extracted. The encoder $E_{1}$ obtains the features as the input and produce latent representations of both clean speech signal and the mixture. In the proposed method, we consider two pre-tasks for pre-training: latent representation and mask estimation. The first task aims to learn the latent representation of only clean speech signals. However, the second task trains DM and ERM to describe the representation relationships from the target speech signal to the mixture. Both the latent representation and masks are trained by minimizing the discrepancy between the input representation and the corresponding reconstruction. The decoder obtains the averaged masked representations from two tasks and produces the estimated speech signal. 

Both $E_{1}$ and $D_{1}$ consist of 4 1-D convolutional layers. In $E_{1}$, the size of the hidden dimension sequentially decreases from 512 $\rightarrow$ 256 $\rightarrow$ 128 $\rightarrow$ 64. Consequently, the dimension of the latent space is set to 64, and a stride of 1 sample with a kernel size of 7 for the convolutions. Different from $E_{1}$, $D_{1}$ increases the size of the latent dimensions inversely.


Different from the PAE, the DAE only needs access to the reverberant mixture. The feature is extracted from the reverberant mixture and input to $E_{2}$. Consequently, the latent representation of the mixture is obtained as the output of $E_{2}$. The learnt representation and masks from the PAE are exploited to modify the loss functions and learn a shared latent space between the clean speech and mixture representations. Benefited from the pre-tasks, a mapping from the mixture domain to the target speech domain is learnt with the latent representation of the clean speech signal. Furthermore, $D_{2}$ is trained to produce the estimated mixture as the downstream task.

The DAE network follows a similar architecture to PAE. $E_{2}$ consists of 6 1-D convolutional layers where the hidden layer sizes decrease from 512 $\rightarrow$ 400 $\rightarrow$ 300 $\rightarrow$ 200 $\rightarrow$ 100 $\rightarrow$ 64, and $D_{2}$ increases the sizes inversely.

In the testing stage, after the features are extracted from the reverberant mixtures, they are fed into the trained $E2$ as the input. As aforementioned, the loss function in $E2$ is trained with the mapping of the latent space from the mixture domain to the target speech domain. Thus, the trained $E2$ produces an estimated latent representation of the reverberant mixture. Then, the estimated masks are used to dereverberate and denoise the mixture representation. Finally, the trained $D1$ obtains the reconstructed representation and maps to the target speech signal. 

\vspace{-0.5em}
\subsection{Masking Module}
As aforementioned, the masking module is exploited to trains DM and ERM to describe the representation relationships from the target speech signal to the mixture. The architecture of the masking module is depicted in Fig. 1 (b).


The masking module has three sub-layers. The encoder produces the mixture representation $\mathbf{Y}$, and the aim of the masking module is to estimate the target speech representation $\hat{\mathbf{S}}$. The first two sub-layers consists of two time-frequency (TF) masks, DM and ERM, respectively. According to \cite{Two}, DM is presented:
\begin{equation}
DM=(\mathbf{S}+\mathbf{I})\cdot\mathbf{Y}^{-1}
\end{equation}
where $`\cdot$' is the dot product, and $\mathbf{S}$ and $\mathbf{I}$ are the clean speech signal and the interference, respectively. The dereverberated mixture is obtained as: 
\begin{equation}
\hat{\mathbf{Y}}_{d}=\mathbf{Y}\cdot \widehat{D M}
\end{equation}
where $\widehat{D M}$ is the estimated DM. However, in practice, obtaining the dereverberated mixtures is very challenging \cite{tf3}. Although most of the reverberations are removed by DM, the remaining reverberations in $\hat{\mathbf{Y}}_{d}$ still limit the performance \cite{IET}. Thus, in the second sub-layer, we exploit ERM in the second sub-layer to further improve the speech enhancement in reverberant environments, which can be defined as:
\begin{equation}
\widehat{ERM}=\frac{{\mid}\mathbf{S}{\mid}}{{\mid}{\hat{\mathbf{Y}}_{d}}{\mid}}
\end{equation}
Then, the background noise and the remaining reverberations are removed by ERM. Moreover, the ReLU activation is added to each mask and produces the output for the next sub-layer. Additionally, a residual connection \cite{rc} is applied in the masking module to ease the training of the module. Finally, the target speech representation is obtained with a PReLU activation \cite{prelu} as:
\begin{equation}
{\hat{\mathbf{S}}}={\widehat{ERM}}\cdot{\widehat{DM}}\cdot{{\mathbf{Y}}}
\end{equation}

The overall loss to train the masking module is a combination of three loss terms as:
\begin{equation}
\mathcal{L}_{\text {masking }}=\theta_{1}  \cdot \mathcal{L}_{\text {KL-masking }}+\mathcal{L}_{\mathbf{S}}+\mathcal{L}_{\text {cyc }}
\end{equation}
where $\mathcal{L}_{\text {KL-masking }}$ denotes the Kullback-Leibler (KL) loss and is applied to train the latent representation closed to a normal distribution \cite{ssl}. Then, $\theta_{1}$ is the coefficient of $\mathcal{L}_{\text {KL-masking }}$ and empirically set to 0.001. Besides, $\mathcal{L}_{\mathbf{S}}$ denotes the loss between the target speech signal and the corresponding reconstruction. The $L$2 norm of the error $\mathcal{L}_{\mathbf{S}}=\|\mathbf{S}-\widehat{\mathbf{S}}\|_{2}^{2}$ is exploited as the loss function. Similarly, the cycle loss $\mathcal{L}_{\mathrm{cyc}}$ consists of $\mathcal{L}_{\mathrm{S}}$ and the loss between the latent representation and the corresponding reconstruction:
\begin{equation}
\mathcal{L}_{\mathrm{cyc}}=\|\mathbf{S}-\widehat{\mathbf{S}}\|_{2}^{2}+\theta_{2} \cdot\left\|\mathbf{X}_{S}-\widehat{\mathbf{X}}_{S}\right\|_{2}^{2}
\end{equation}
where $\widehat{\mathbf{X}}_{\mathbf{S}}$ is the estimated representation of the target speech signal. Moreover, $\theta_{2}$ is the coefficient of representation loss and empirically set to 0.001. Finally, the combination of losses are utilized in PAE to improve the speech enhancement performance.

\vspace{-0.5em}
\section{EXPERIMENTAL RESULTS}
\label{sec:pagestyle}
\subsection{Experimental Setup}
The proposed method is trained by using the Adam optimizer with a learning rate of 0.001 and the batch size is 20. The number of epochs for PAE and DAE are 700 and 1500, respectively. All the experiments are run on a work station with four Nvidia GTX 1080 GPUs and 16 GB of RAM. The magnitude spectrograms have 513 frequency bins for each frame as a Hanning window and a discrete Fourier transform (DFT) size of 1024 samples are applied.



To evaluate the proposed model, we use composite metrics that approximate the Mean Opinion Score (MOS) including COVL: MOS predictor of overall signal quality, CBAK: MOS predictor of background-noise intrusiveness, CSIG: MOS predictor of signal distortion \cite{mos} and Perceptual Evaluation of Speech Quality (PESQ). Higher values of the measurements imply that the desired speech signal is better estimated.

\vspace{-1em}
\subsection{Comparisons and Datasets}
We compare the proposed method with two recent SSL speech enhancement approaches \cite{ssl}\cite{ssl1}. The first one is SSE \cite{ssl} which exploits two autoencoders to process pre-task and downstream task, respectively. The architecture is similar to the proposed method. The second one is pre-training fine-tune (PT-FT) \cite{ssl1}, which uses three models and three SSL approaches for pre-training: speech enhancement, masked acoustic model with alteration (MAMA) used in TERA \cite{mama} and continuous contrastive task (CC) used in wav2vec 2.0 \cite{wav}. We reproduce the PT-FT method with DPTNet model \cite{DT} and speech enhancement as the pre-task because it shows the best enhancement performance in \cite{ssl1}. 

\vspace{-1em}
\begin{table}[htbp!]
\caption{Comparison of SSL speech enhancement approaches with the proposed method. More specifically, the PT-FT method use 50,800 paired utterances in the training stage. However, only 200 utterances are required in the proposed method. Besides, 3 pre-tasks are trained in the PT-FT method and we train 2 pre-tasks in the proposed method.}
\vspace{1em}
\centering
\begin{tabular}{c|c|c|c}
\hline
 & SSE \cite{ssl}  & PT-FT \cite{ssl1} &\textit{Proposed } \\    
 \hline
Noise & \xmark & \checkmark  & \checkmark    \\
\hline
Paired Data & \xmark & \checkmark & \checkmark    \\
\hline
  Multiple Models   &\checkmark&\xmark & \checkmark\\
 \hline 
  Single Pre-Task  &\checkmark &\xmark & \xmark\\
 \hline 
  Reverberation  &\checkmark &\xmark &\checkmark\\
 \hline 
\end{tabular}
\end{table}

\vspace{-0.5em}
To evaluate the speech enhancement performance, in the training stage, 600 clean utterances from 20 speakers with three room environments are randomly selected from the DAPS dataset \cite{daps}. The training data consists of 10 male and 10 female speakers each reading out 5 utterances and recorded in different indoor environments with different real room impulse responses (RIRs). In each environment, we first randomly select 12 utterances to generate the pair of training data as clean speech signals and mixtures to train the PAE. Then, the rest 188 utterances are exploited for DAE to obtain the estimated mixtures. Moreover, we use three background noises ($factory$, $babble$, and $cafe$) from the NOISEX dataset \cite{noise} and three SNR levels (-5. 0, and 5 dB) to generate the mixtures. In the testing stage, 300 clean utterances of 10 speakers are randomly selected and used to generate the mixtures with the same background noises and SNR levels as the configuration in training stage.

\vspace{-0.5em}
\subsection{Results and Discussions}
In the evaluations, we first conduct the experiments in three cases as different interferences in the PAE:

$\bullet $ Case 1: The interference only consists of the background noises ($factory$, $babble$, and $cafe$).

$\bullet $ Case 2: In the SSE method shown in \cite{ssl}, only limited amount of clean speech signals and unlabeled mixtures are available in the training stage. Therefore, to further evaluate the proposed masking module, we randomly generate a Gaussian noise to produce the reverberant mixture as the interference. Hence, compared with \cite{ssl}, no extra information is introduced.


$\bullet $ Case 3: To evaluate the performance with various interferences, we use both the background noise (Case 1) and the unlabelled mixture (Case 2) to generate the interference. In both Cases 2 $\&$ 3, the mixtures used in the PAE and the DAE are unseen to the each other.

\vspace{-0.5em}
\subsubsection{Case 1}
\vspace{-2em}

\begin{table}[htbp!]
\caption{Averaged speech enhancement performance (Case 1) in terms of three room environments, three noise interferences and three SNR levels.}
\vspace{1em}
\centering
\begin{tabular}{ccccc}
\hline
Method & PESQ  & CSIG &CBAK &COVL\\    
 \hline
SSE \cite{ssl} &1.48 & 2.28 & 1.90&1.84  \\
PT-FT \cite{ssl1} &1.58 & 2.34 & 2.04&1.91  \\
  \textit{Proposed }  &{\bfseries 1.71} &{\bfseries 2.45} & {\bfseries 2.16}& {\bfseries 1.97}\\
 \hline 
\end{tabular}
\end{table}

From Table 2, it is clearly observed that the proposed method outperforms the state-of-the-art methods in terms of all three performance measurements. In \cite{ssl1}, the original PT-FT method is trained with Libri1Mix train-360 set \cite{lib} which contains 50,800 utterances. However, in the comparison experiments, we use the limited amount of training utterances (200). Therefore, the speech enhancement performance of the PT-FT suffers a significant degradation compared with the original paper. The latent representation and the masking module have limitations, however, the proposed method takes advantage of both approaches and mitigates the speech enhancement problem. Thus, the speech enhancement performance is improved compared with only learning the clean speech representation in the SSE method.
\vspace{-0.5em}
\subsubsection{Case 2}

\vspace{-2em}
\begin{table}[htbp!]
\caption{Averaged speech enhancement performance (Case 2) in terms of three room environments, three noise interferences and three SNR levels.}
 \vspace{1em}
\centering
\begin{tabular}{ccccc}
\hline
Method & PESQ  & CSIG &CBAK &COVL\\    
 \hline
SSE \cite{ssl} &1.39 & 2.31 & 1.82&1.75  \\
PT-FT \cite{ssl1}  &1.44 & 2.34 & 1.89&1.90  \\
  \textit{Proposed} & {\bfseries 1.64} & {\bfseries 2.38} &  {\bfseries 2.11}&{\bfseries 1.92}\\
 \hline 
\end{tabular}
\end{table}

It can be seen from Table 3 that the proposed method always achieves highest enhancement performance compared with SSE and PT-FT. However, the enhancement performance at Case 2 suffers a degradation compared with Case 1. Because in Case 2, the interference consists of the undesired speech signal, the background noise, and reverberation of both speech signals and noises. It is highlighted that, due to different distributions between speech and noise interference domains, the task of personalized speech enhancement from the mixture with undesired speech signals is more challenging than from noise interferences \cite{dav}.

\subsubsection{Case 3}
In this case, we use two background noises ($restaurant$ and $f16$) from the NOISEX dataset \cite{noise} and two SNR levels (-5 and 5 dB). The experimental results are shown in Table 4.

\vspace{-0.5em}
\begin{table}[htbp!]
\caption{Averaged speech enhancement performance (Case 3) in terms of three room environments, two noise interferences and two SNR levels.}
 \vspace{1em}
\centering
\begin{tabular}{ccccc}
\hline
Method & PESQ  & CSIG &CBAK &COVL\\    
 \hline
SSE \cite{ssl}& 1.37& 2.27 & 1.77 &1.66  \\
PT-FT \cite{ssl1}  &1.49 & 2.25 & 1.84&1.87  \\
\textit{Proposed }  &{\bfseries 1.69} &{\bfseries 2.29} & {\bfseries 2.13}& {\bfseries 1.90}\\
 \hline 
\end{tabular}
\end{table}

It can be observed from Table 4 that the speech enhancement performance is significantly improved by the proposed method compared to the baselines. Although Table 3 indicates the case where the interference consists of the background noise and the reverberant mixture. The improvement in terms of PESQ, CBAK, and COVL is more obvious than the other two cases.

In all comparison experiments, we can observe that: (1) The proposed method outperforms the recent SSL-based speech enhancement methods. (2) When the interference contains both background noise and the undesired speech signal, the enhancement performance is degraded. (3) The proposed method still improves the speech enhancement performance in the hardest case (Case 3) because unseen scenario is also considered. Moreover, the improvement is more significant as the case is more challenging.
\section{Conclusion}
\label{sec:prior}

In order to address the monaural speech enhancement problem in reverberant environments, a multi pre-tasks SSL method was proposed. In the pre-training stage, the latent representation of the clean speech signal was learnt as the first pre-task. Meanwhile, in the PAE, a DM- and ERM-based masking module was applied to assist to estimate the target speech representation. We evaluated the proposed method in three cases with different interferences. The experimental results showed that pre-training with multi pre-tasks provides better speech enhancement performance than the state-of-the-art approaches within the benchmark dataset. 


\vfill\pagebreak





\bibliographystyle{IEEEbib}
\bibliography{Template}

\end{document}